\documentclass[12pt]{iopart}
\usepackage{iopams}
\usepackage{graphicx} 
\usepackage{color}
\usepackage{bm}

\begin{document}

\newcommand \be {\begin{equation}}
\newcommand \ee {\end{equation}}
\newcommand \bea {\begin{eqnarray}}
\newcommand \eea {\end{eqnarray}}

\newcommand \figref[1] {Fig.~\ref{#1}}

\newcommand{\ecut}{\epsilon^{\dagger}}
\newcommand{\limInf}[1]{\lim_{#1 \rightarrow + \infty} }
\newcommand{\Eqref}[1]{Eq.~(\ref{#1})}
\newcommand{\Esp}[1]{\left< #1 \right>}
\newcommand{\Prob}[1]{\mathbb{P}\p{#1}}
\newcommand{\p}[1]{\left(#1\right)}
\newcommand{\R}{\mathbb{R}}

\newcommand{\Comment}[1]{}

\title[]{On the existence of a glass transition in a Random Energy Model}

\author{Florian Angeletti, Eric Bertin, Patrice Abry}

\address{Universit\'e de Lyon, Laboratoire de Physique,
\'Ecole Normale Sup\'erieure de Lyon, CNRS,
46 all\'ee d'Italie, F-69007 Lyon, France
}
\ead{florian.angeletti@ens-lyon.fr, eric.bertin@ens-lyon.fr, patrice.abry@ens-lyon.fr}

\begin{abstract}
We consider a generalized version of the Random Energy Model in which the energy of each configuration is given by the sum of $N$ independent contributions (``local energies'') with finite variances but otherwise arbitrary statistics. Using the large deviation formalism, we find that the glass transition generically exists when local energies have a smooth distribution. 
In contrast, if the distribution of the local energies has a {Dirac mass} at the minimal energy (e.g., if local energies take discrete values), the glass transition ceases to exist if the number of energy levels grows sufficiently fast with system size.
This shows that statistical independence of energy levels does not imply the existence of a glass transition.
\end{abstract}


\section{Introduction}

In the context of condensed matter physics, the glass transition
is a rather generic phenomenon through which
the state of a system becomes partially frozen below a threshold temperature
\cite{Berthier,Nagel,book-glasses}.
One of the simplest models exhibiting a glass transition
is the Random Energy Model (REM) introduced by Derrida
\cite{Derrida},
in which the energies of microscopic configurations are independent
and identically distributed random variables drawn from a distribution
$\mathcal{P}(E)$,
chosen to be Gaussian in the original version of the model.
The REM provides a simple illustration of the so-called
'one-step replica symmetry breaking' scenario of the glass transition
\cite{BouchMez97}, which is known to hold in more sophisticated mean-field
models \cite{Wolynes87,Wolynes89}.
Several generalizations of the REM have been proposed, mostly to
incorporate correlations in simple ways
\cite{Gardner,Carpentier,Fyodorov,Rosso}.
In addition, interesting connections of the REM with probabilistic issues
such as the convergence properties of sums \cite{BenArous} and extreme values
\cite{Bogachev,Fyodorov,AngelettiEVS} of random variables, as well as
with signal processing issues such as moment estimation
\cite{Angeletti11}, have been pointed out.
Potential connections with string theories have even been recently
outlined \cite{Saakian09}.

In this note, we explore the question whether uncorrelated random
energy levels are enough to generate a glass transition.
To this aim, we consider a generalized version of the REM
offering some freedom in the energy distribution as well as
in the scaling of the number of configuration with system size.
The paper is organized as follows. The model is introduced in
Sect.~\ref{sect-model}, and the necessary framework to analyze
the glass transition is presented in Sect.~\ref{sect-glass-trans}.
Then in Sect.~\ref{sect-no-glass}, which constitutes the core of this
paper, we derive necessary conditions for the absence of glass transition,
and show that these conditions are also sufficient. In addition,
we determine the behavior of the glass transition temperature close to onset.
Finally, Sect.~\ref{sect-discus} summarizes the results,
and discusses some analogies with related problems.

\section{Model}
\label{sect-model}

\subsection{Definition}

We consider a disordered system having a number $M$ of microscopic
configurations, labeled by index $k=1,\ldots,M$; to each configuration
is associated a quenched random energy $E_k$.
The energies $E_k$ are assumed to be independent and identically distributed.
At variance with the standard REM, we do not directly specify the distribution
$\mathcal{P}(E)$ from which the energies $E_k$ are drawn, but we rather assume
that $E_k$ is given by a sum of $N$ individual contributions,
\begin{equation} \label{eq:Es-sum}
E_k = \sum_{i=1}^{N} \eta_{k,i}, \qquad k=1,\ldots,M,
\end{equation}
where the terms $\eta_{k,i}$ are independent and identically distributed
random variables with a finite variance distribution $p(\eta)$.
A standard assumption is that the number $M$ of configurations scales as
\begin{equation} \label{eq:scal-M}
M \sim e^{\alpha N} \qquad (\alpha >0).
\end{equation}
For the standard REM, $\alpha=\ln 2$.
Note that this model may be considered as a simplified version of
the directed polymer problem in a random media \cite{Spohn},
neglecting correlations between local energies on different paths.

One can interpret \Eqref{eq:Es-sum} as the decomposition of the total
energy over the $N$ degrees of freedom of the system, with the strong
assumption that local contributions associated to different microscopic
configurations $k$ are statistically independent.
Such an approach provides an alternative
interpretation (besides the standard one in terms of $p$-spin model
in the limit $p \to \infty$ \cite{Derrida}) of the Gaussian energy
distribution of variance proportional to $N$ used in the standard
REM \cite{Derrida} (though more general energy distributions have
also been considered \cite{BouchMez97}).
But it also gives many ways to depart from the Gaussian distribution,
in the sense that equilibrium at finite temperature is dominated
by energy values in the lower tail of the energy distribution,
far from the maximum of the distribution around which the Gaussian
approximation holds.
As we shall see below, the key ingredient of our generalized version of REM
is the introduction of both the arbitrary distribution $p(\eta)$ and
the free parameter $\alpha$, which leads to a richer behavior than
in the standard REM.

\subsection{Large deviation function}

To characterize the distribution $\mathcal{P}(E)$ of the energies $E_k$ from the distribution $p(\eta)$,
we use large deviation theory \cite{Ellis,Touchette}.
Let us define the energy density $\epsilon=E/N$, and its distribution
$P(\epsilon) = N \mathcal{P}(N\epsilon)$.
The G\"artner-Ellis theorem \cite{Ellis,Touchette}
implies that, if the cumulant generating function
\begin{equation} \label{def-lambda}
\lambda(q) = \ln \Esp{e^{q \eta}}
\end{equation}
 is defined and differentiable on the real axis, then the distribution $P(\epsilon)$ is given by
\begin{equation}
\label{eq@LDf}
P(\epsilon) \approx e^{- N \phi(\epsilon) }
\end{equation}
where $\phi$ is the Legendre transform of $\lambda$,
$\phi(\epsilon) = \max_{q} \{\epsilon q - \lambda(q)\}$.
Note that the properties of the Legendre transform imply that $\phi(\epsilon)$
is a convex function. Similarly, since for all $\epsilon$, $(\epsilon q - \lambda(q))_{|q=0} = 0$, we have $\phi(\epsilon) \ge 0
$. Moreover, the lower bound $0$ is attained at $\epsilon=\lambda'(0)$: $\phi(\lambda'(0))=0$.
In the present statistical physics context, it is however more natural
to use the function 
\begin{equation} \label{def-mu}
\mu(\beta) = \ln \Esp{e^{-\beta \eta}} =\lambda(-\beta)
\end{equation}
instead of $\lambda(q)$, and we shall thus use $\mu(\beta)$ in the following,
together with the relation
\begin{equation} \label{eq:legendre-mu}
\phi(\epsilon) = -\min_{\beta} \{\epsilon \beta + \mu(\beta)\}.
\end{equation}

\section{Analysis of the glass transition}
\label{sect-glass-trans}

In the present section, we briefly set up the general framework
allowing the glass transition to be studied. This framework
mainly relies on the existence of a finite size cutoff in the
density of state \cite{Derrida}.

\subsection{Finite size cutoff}
\label{sec@Fsc}

Considering a given sample with a finite number of configurations,
the energies $E_k$ are necessarily
confined to a finite subdomain of the support of $\mathcal{P}(E)$.
In the low temperature regime, the lower bound of this domain is
known to play an important role \cite{Derrida}.
Although the boundaries of this domain are also random variables,
it is nevertheless possible to define a sharp lower boundary on the energy
density in the large $M$ limit.
Defining the cumulative $F(\epsilon) = \int_{-\infty}^{\epsilon} d\epsilon' \, P(\epsilon')$,
we consider the probability
\begin{equation} \label{eq:pe1eM}
\Prob{\epsilon_1,\dots,\epsilon_M > \epsilon} = [1-F(\epsilon)]^M
\end{equation}
that all energy densities $\epsilon_k$, $k=1,\ldots,M$, are larger
than a given value $\epsilon$.
We denote as $\epsilon_0$ the value of $\epsilon$ for which $P(\epsilon)$
is maximum (i.e., $\phi(\epsilon_0)=0$).
Note that $\phi(\epsilon)$, being a convex function, necessarily
decreases for $\epsilon<\epsilon_0$.
Using a standard saddle-point method and neglecting non-exponential
prefactors, we have for $\epsilon<\epsilon_0$,
$F(\epsilon) \approx e^{- N \phi(\epsilon)}$ so that
$[1-F(\epsilon)]^M \approx e^{-M F(\epsilon)}$.
\Eqref{eq:pe1eM} can then be rewritten as
\begin{equation} \label{eq:pe1eM2}
\ln \Prob{\epsilon_1,\dots,\epsilon_M > \epsilon} \approx -M\,F(\epsilon)
= - e^{N(\alpha-\phi(\epsilon))},
\end{equation}
using $\ln M = \alpha N$ [see \Eqref{eq:scal-M}].
Let us define $\ecut<\epsilon_0$ such that
\begin{equation} \label{eq@ecut}
\phi( \ecut )  = \alpha.
\end{equation}
\Eqref{eq:pe1eM2} shows that $\Prob{\epsilon_1,\dots,\epsilon_M > \epsilon}$
exhibits a sharp crossover at $\epsilon=\ecut$.
For $\epsilon < \ecut$, $\alpha-\phi(\epsilon)<0$,
and $\Prob{\epsilon_1,\dots,\epsilon_M > \epsilon} \to 1$ when $N \to \infty$,
so that there is with probability one no energy levels below $\epsilon$.
In contrast, for $\epsilon > \ecut$, $\alpha-\phi(\epsilon)>0$, so that
$\Prob{\epsilon_1,\dots,\epsilon_M > \epsilon} \to 0$ when $N \to \infty$,
which means that with probability one, there are energy levels lower than
$\epsilon$. In the limit $N \to \infty$, the value $\ecut$ thus corresponds
essentially to a border below which there are no more energy levels.
In other words, the value $\ecut$ can be considered as the ground state energy
density in almost all samples
\footnote{Note that similarly, an upper bound also exists for the energy levels,
but we do not take it into account as it plays no role in the thermodynamic
properties.}.

\subsection{Free energy and glass transition temperature}

To evaluate the disorder averaged free energy $\Esp{F}=-\beta^{-1}\Esp{\ln Z}$,
where $Z$ is the partition function $Z = \sum_{k=1}^M e^{-\beta E_k}$
and $\beta=1/T$ the inverse temperature,
a usual method (beyond the replica trick \cite{BouchMez97}) is
to determine the typical value
$Z_{\mathrm{typ}}$ (rather than the averaged one) of the partition function,
yielding $\Esp{F} \approx -T \ln Z_{\mathrm{typ}}$.
$Z_{\mathrm{typ}}$ is evaluated taking into account the threshold $\ecut$, and
approximating the density of states $n(\epsilon)$
by the disorder averaged one $\Esp{n(\epsilon)}=M\, P(\epsilon)$ in the energy
range $\epsilon > \ecut$, where $\Esp{n(\epsilon)}$ is large.
One finds, using the large deviation form \Eqref{eq@LDf},
\begin{equation} \label{eq:ztyp}
Z_{\mathrm{typ}} = \int_{\ecut}^{\infty}  e^{N g(\epsilon)} d\epsilon,
\qquad g(\epsilon) = \alpha-\phi(\epsilon) -\beta \epsilon.
\end{equation}
For large $N$, the partition function \Eqref{eq:ztyp} can be evaluated
through a saddle-point approximation.
Equating the derivative $g'(\epsilon)$ to zero yields
\begin{equation} \label{eq:phiprime-beta}
\phi'(\epsilon_{\mathrm{m}})=-\beta.
\end{equation}
As $g''(\epsilon) \le 0$, $g(\epsilon)$ decreases for
$\epsilon > \epsilon_{\mathrm{m}}$.
If $\epsilon_{\mathrm{m}} > \ecut$, the saddle-point evaluation yields
$Z_{\mathrm{typ}} \approx e^{N(\alpha-\phi(\epsilon_{\mathrm{m}}) -\beta \epsilon_{\mathrm{m}})}$.
In the opposite case $\epsilon_{\mathrm{m}} < \ecut$, the global maximum of
$g(\epsilon)$ is no longer relevant as it falls outside the integration
interval. The maximum of $g(\epsilon)$ over the interval $[\ecut,\infty)$
is then $g(\ecut)$, leading to
$Z_{\mathrm{typ}} \approx e^{N(\alpha-\phi(\ecut) -\beta \ecut)}$.
The border between these two regimes, $\epsilon_{\mathrm{m}} = \ecut$,
defines the glass transition temperature $T_g\equiv \beta_{\mathrm{g}}^{-1}$
through the implicit relation
\begin{equation} \label{eq:epsm-ecut}
\epsilon_{\mathrm{m}}(\beta_{\mathrm{g}}) = \ecut.
\end{equation}
As $\phi(\epsilon)$ is a convex function, $\phi'(\epsilon)$ is an increasing
function of $\epsilon$. Hence from \Eqref{eq:phiprime-beta},
$\epsilon_{\mathrm{m}}$ is a decreasing function of $\beta$.
The glassy regime $\epsilon_{\mathrm{m}} < \ecut$ thus corresponds to $\beta>\beta_{\mathrm{g}}$,
or equivalently to $T<T_g$.
Altogether, the free energy per degree of freedom $f=\Esp{F}/N$ reads
for $N \to \infty$, using \Eqref{eq@ecut},
\bea
f(\beta) &=& \frac{1}{\beta}\, \phi\big(\epsilon_{\mathrm{m}}(\beta)\big) + \epsilon_{\mathrm{m}}(\beta)
- \frac{\alpha}{\beta}, \qquad \beta < \beta_{\mathrm{g}}, \\
f(\beta) &=& \ecut, \qquad \beta > \beta_{\mathrm{g}}.
\eea
Inverting the Legendre transform \Eqref{eq:legendre-mu}, one finds
\begin{equation} \label{eq:mu-beta-phi}
\mu(\beta) = -\beta \epsilon_{\mathrm{m}}(\beta) - \phi\big(\epsilon_{\mathrm{m}}(\beta)\big)
\end{equation}
so that the free energy can be rewritten for $\beta < \beta_{\mathrm{g}}$ as
$f(\beta) = -\frac{1}{\beta}\, \mu(\beta) - \frac{\alpha}{\beta}$.
Introducing the function
\begin{equation} \label{eq:def-zeta}
\zeta(\beta) = \beta \mu'(\beta) - \mu(\beta),
\end{equation}
the entropy $s=-df/dT=\beta^2 df/d\beta$ then reads\Comment{
\begin{equation} \label{eq-entropy}
s(\beta) = \begin{cases}
\alpha - \zeta(\beta)&  \beta < \beta_{\mathrm{g}}, \\
  0 & \beta > \beta_{\mathrm{g}}.
\end{cases}
\end{equation}}
\begin{equation} \label{eq-entropy}
s(\beta) = \left\{ \begin{array}{rl}
\alpha - \zeta(\beta),&  \quad \beta < \beta_{\mathrm{g}}, \\
  0, & \quad \beta > \beta_{\mathrm{g}}.
\end{array} \right.
\end{equation}
Note that $\zeta'(\beta)=\beta \mu''(\beta)$ and that $\mu(\beta)$ is a
convex function with $\mu(0)=0$, so that $\zeta(\beta)$ is an increasing function of $\beta$,
starting from $\zeta(0)=0$.
Accordingly, the entropy $s$ is a decreasing function of $\beta$
for $\beta < \beta_{\mathrm{g}}$.
We thus recover in this general framework the standard interpretation
of the glass transition in terms of a vanishing entropy per degree
of freedom, meaning
that in the low temperature phase, the probability distribution
concentrates on a few microscopic configurations \cite{Derrida}.
\Eqref{eq-entropy} provides us with an alternative characterization
of the transition temperature $\beta_{\mathrm{g}}$
\begin{equation} \label{eq-betag-alt}
\zeta(\beta_{\mathrm{g}}) = \alpha.
\end{equation}

From \Eqref{eq-betag-alt}, one sees that the glass transition exists
if the function $\zeta(\beta)$, which is defined for all $\beta>0$, reaches
the value $\alpha$ for some finite inverse temperature $\beta_{\mathrm{g}}$.
In the standard REM, which is recovered by choosing for $p(\eta)$
a centered Gaussian distribution of variance $J^2/2$,
one has $\zeta(\beta)=J^2 \beta^2/4$, so that $\zeta(\beta)$ can reach
any value $\alpha$, implying the existence of the glass transition.
More precisely, as shown in the left panel of \figref{fig:frozen}, we have
\begin{equation} \label{eq:betag-gauss}
\beta_{\mathrm{g}}(\alpha) = \sqrt{\frac{4 \alpha}{J^2} }.
\end{equation}
However, in the present more general setting of an arbitrary $p(\eta)$,
$\zeta(\beta)$ may be bounded and
the glass transition may not exist, as we shall see in the next section.

\section{Conditions for the existence or absence of glass transition}
\label{sect-no-glass}

\begin{figure}
\centerline{ \includegraphics[width=5cm]{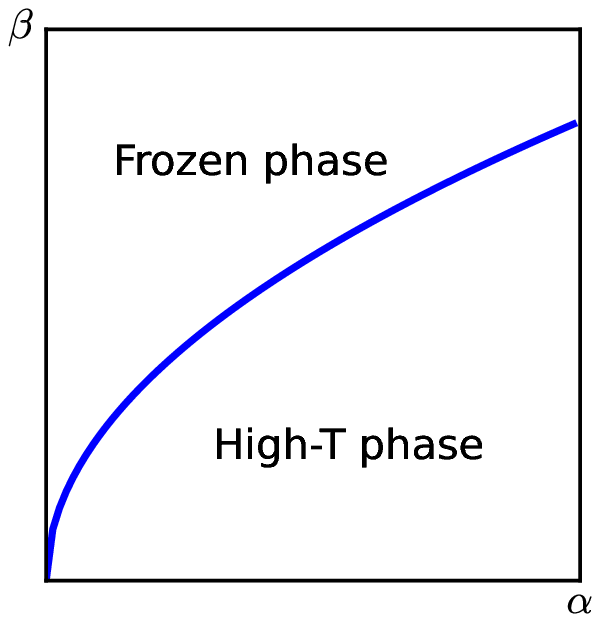} \includegraphics[width=5cm]{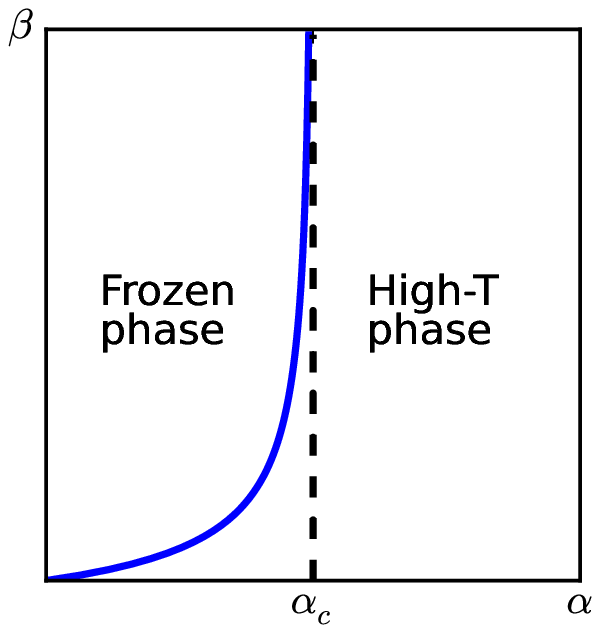} }
\caption{\label{fig:frozen} Sketch of the phase diagram of the REM in the $(\alpha,\beta)$ plane.
Left: case when the function $\zeta(\beta)$ diverges when $\beta$ goes to infinity. The line corresponds to the glass transition (inverse) temperature $\beta_{\mathrm{g}}(\alpha)$, illustrated here on
the standard REM with a Gaussian energy distribution --see \Eqref{eq:betag-gauss}.
Right: same diagram in the case when the function $\zeta(\beta)$ has a finite limit $\alpha_{\mathrm{c}}$. The vertical dashed line $\alpha=\alpha_{\mathrm{c}}$ corresponds to the asymptote of $\beta_{\mathrm{g}}(\alpha)$.
For $\alpha>\alpha_{\mathrm{c}}$, no glass transition occurs.}
\end{figure}

In this section, we study the asymptotic behavior of $\zeta(\beta)$ for
$\beta \to \infty$, to see if $\zeta(\beta)$ converges to a finite limit,
or diverges.
If $\zeta(\beta)$ diverges when $\beta \to \infty$, it will necessarily
cross (assuming continuity) the value $\alpha$ for some finite $\beta_{\mathrm{g}}$.
In constrast, if $\zeta(\beta)$ converges to a finite limit $\zeta_{\infty}$,
\Eqref{eq-betag-alt} has a solution only if $\alpha<\zeta_{\infty}$. Hence
a glass transition exists only if $\alpha < \alpha_{\mathrm{c}} \equiv \zeta_{\infty}$
as illustrated on the right panel of \figref{fig:frozen}.
The question is then to know for which form of the distribution $p(\eta)$
--see \Eqref{def-mu}-- the function $\zeta(\beta)$ can have a finite limit $\alpha_{\mathrm{c}}$.

\subsection{Necessary conditions for the absence of glass transition}

In this section, we wish to derive some necessary conditions for the absence
of glass transition. We thus start by assuming that $\zeta(\beta)$ has a finite limit
$\alpha_{\mathrm{c}}$ and we explore the implications of this assumption on the distribution $p(\eta)$
of the local energies.

\subsubsection{Asymptotic behavior of $\mu'(\beta)$ and $\mu(\beta)$\\}

In order to derive the behavior of $\mu(\beta)$ from the hypothesis that $\zeta(\beta)$ admits a finite limit, it is useful to express $\mu(\beta)$ as a functional of $\zeta(\beta)$ by solving \Eqref{eq:def-zeta} as a differential equation in $\mu$, for a given function $\zeta$. 
Using classical ordinary differential equation method, the following result is obtained
\begin{equation} \label{eqn-mu[zeta]}
\mu(\beta)= - \beta \left[ \int_{\beta}^{+\infty} \frac{\zeta(t)} {t^2} dt - \mu'_{\infty} \right],
\end{equation}
where $\mu'_{\infty}$ is a constant.
One should note that the integral in \Eqref{eqn-mu[zeta]} is always well-defined if $\zeta$ admits a finite limit.

The next step is to use this integral form of $\mu$ to show that 
 $\mu'_{\infty}$ is rightfully the limit of $\mu'(\beta)$ when  $\beta \to +\infty$.
Differentiating \Eqref{eqn-mu[zeta]} yields
\begin{equation} \label{eqn-mu'[zeta]}
\mu'(\beta)=  - \left[ \int_{\beta}^{+\infty} \frac{\zeta(t)} {t^2} dt - \mu'_{\infty} \right]
+ \frac{\zeta(\beta)}{\beta}.
\end{equation}  
Consequently, in the limit $\beta \rightarrow + \infty$, $\mu'(\beta)$ admits a finite limit 
\begin{equation} \label{eqn-mu'[zeta]-lim}
\limInf{\beta} \mu'(\beta)=  \mu'_{\infty}.
\end{equation}  
Note that from the convexity of $\mu(\beta)$, $\mu'(\beta)$ is an increasing
function, so that
\begin{equation} \label{eq:mu'-muinfty}
\mu'(\beta) -  \mu'_{\infty} \le 0,
\end{equation}
a property which will prove useful later on.
We shall show that $\mu(\beta)$ has a linear asymptote, which is actually
not obvious from \Eqref{eqn-mu'[zeta]-lim}
\footnote{For instance, a function of the type $f(x)=ax+b\ln x$ has a finite
derivative equal to $a$ when $x \to +\infty$, but has no linear asymptote.}.
\Eqref{eqn-mu[zeta]} can be rewritten as
\begin{equation} \label{eqn-mu[zeta]:0}
\mu(\beta)- \beta \mu'_{\infty} = - \beta \int_{\beta}^{+\infty} \frac{\zeta(t)} {t^2} dt.
\end{equation}
Since $\zeta(\beta)$ admits a finite upper bound $K$, we have
 
\begin{equation}  \label{eq-linasympt-bound}
\mu(\beta) - \beta \mu'_{\infty} \ge - \beta \int_{\beta}^{+\infty} \frac{K} {t^2} dt  = -K.
\end{equation}
From \Eqref{eq:mu'-muinfty}, $\mu(\beta) - \beta \mu'_{\infty}$ is a decreasing function. 
Therefore, the existence of the lower bound derived in \Eqref{eq-linasympt-bound} implies that $\mu(\beta) - \beta \mu'_{\infty}$ has a finite limit, namely

\begin{equation} \label{eq-mu(infty)}
\limInf{\beta} \mu(\beta) - \beta \mu'_{\infty} = \mu_{\infty}.
\end{equation}
In other words, $\mu(\beta)$ has a linear asymptote with slope $\mu'_{\infty}$.
Furthermore, one necessarily has $\mu_{\infty} \le 0$ since
$\mu(\beta) - \beta \mu'_{\infty}$ is a decreasing function starting from the
value $0$ at $\beta=0$.

\subsubsection{Support of $p(\eta)$\\}

We shall now focus on the case $\mu'_{\infty}=0$, as the case of an arbitrary
value $\mu'_{\infty}$ can be obtained from the case $\mu'_{\infty}=0$
by a shift of the variable $\eta$.
We shall first specify the support of $p(\eta)$.
One of the properties of the cumulant generating function $\lambda$
introduced in \Eqref{def-lambda} is that the image of $\R$ by $\lambda'$
is the support of the probability density function $p(\eta)$, i.e. 
\begin{equation}
\{ \lambda'(\beta),\, \beta \in \R \} = \{ \eta,\, p(\eta) > 0,\, \eta \in \R \}.
\end{equation}
Given that $\mu(\beta)=\lambda(-\beta)$ [see \Eqref{def-mu}],
the lower bound of the support of $p(\eta)$ is directly related to $\mu'_{\infty}$
\begin{equation}
\eta_{\min} \equiv \inf\{\eta,\, p(\eta)> 0\} = - \mu'_{\infty}.
\end{equation}
Notably, $\mu'_{\infty}=0$ implies that only the positive values of $\eta$ have a non-zero probability.
Taking into account the fact that the support of $p(\eta)$ is $[0,\infty)$,
$\mu(\beta)$ can be rewritten as
\begin{equation} \label{eq-mu[p]-restricted}
\mu(\beta) = \ln \int_{0}^{\infty} p(\eta) e^{- \beta \eta} d\eta.
\end{equation}

\subsubsection{Characterization of $p(\eta)$ in the neighborhood of $\eta_{\min}$\\}
\Eqref{eq-mu[p]-restricted} implies that the asymptotic behavior of $\mu(\beta)$ (for $\beta \to \infty$) is directly linked to the behavior of $p(\eta)$
in the neighborhood of $\eta_{\min}=-\mu'_{\infty}$. Moreover, as stated by \Eqref{eq-mu(infty)}, for $\mu'_{\infty}=0$, $\mu(\beta)$ admits a finite limit $\mu_{\infty}$. We shall now explore the consequences on $p(\eta)$ of the existence of this finite limit $\mu_{\infty}$.
 
If $\mu_{\infty}=0$, the property $\mu(0)=0$ and the monotonicity of $\mu(\beta)$ imply that $\mu(\beta) = 0$.
The only possibility for $p(\eta)$ is the degenerate distribution
$p(\eta) =\delta(\eta)$.
This is obviously not a situation of interest, since all configurations
of the model would have the same energy, equal to zero.

If $\mu_{\infty}<0$, $p(\eta)$ has to satisfy 
\begin{equation} \label{eq-LTp-eta}
\limInf{\beta}  \int_0^{\infty} p(\eta) e^{- \beta \eta} d\eta = e^{\mu_{\infty}},
\qquad 0 < e^{\mu_{\infty}}<1 .
\end{equation}
The integral in \Eqref{eq-LTp-eta} can be rewritten as,
assuming $\beta$ has been made dimensionless,
\begin{equation} \label{eq-integ-split}
\int_0^{\infty} p(\eta) e^{- \beta \eta} d\eta =
\int_0^{1/\sqrt{\beta}} p(\eta) e^{- \beta \eta} d\eta +
\int_{1/\sqrt{\beta}}^{\infty} p(\eta) e^{- \beta \eta} d\eta.
\end{equation}
We first note that the second integral in the r.h.s.~of \Eqref{eq-integ-split}
can be bounded as
\begin{equation}
\int_{1/\sqrt{\beta}}^{\infty} p(\eta) e^{- \beta \eta} d\eta \le
e^{-\sqrt{\beta}} \int_{1/\sqrt{\beta}}^{\infty} p(\eta) d\eta \le e^{-\sqrt{\beta}},
\end{equation}
and thus goes to zero when $\beta \to \infty$ for any distribution $p(\eta)$.
We now focus on the behavior of the first integral in the r.h.s.~of \Eqref{eq-integ-split}.
If $p(\eta)$ does not contain a {Dirac mass} (i.e., a Dirac delta) at $\eta=0$, this integral goes to zero when $\beta \to \infty$, while in the presence
of a {Dirac mass} at $\eta=0$, the integral takes a finite limit
when $\beta \to \infty$.
As a result, \Eqref{eq-LTp-eta}, which is a consequence of the assumption
that $\zeta(\beta)$ has a finite limit when $\beta \to \infty$, can only
be satisfied if $p(\eta)$ contains a {Dirac mass} at $\eta=0$
(we recall that $\eta=0$ is the lowest accessible value of $\eta$).
In this latter case, a natural form for $p(\eta)$ then consists of the following mixture:
\begin{equation} \label{eq-p-eta-delta}
p(\eta) = D\, \delta(\eta) + (1-D)\, c(\eta)
\end{equation}
where $0< D<1$ and $c(\eta)$ is a probability density function with support $[0,\infty)$ 
and no Dirac mass at $\eta=0$. For instance, $c(\eta)$ can be
a regular distribution (which may have an integrable divergence at $\eta=0$)
or a sum of Dirac masses, meaning that $\eta$ takes discrete values.

Finally, as mentioned above, the case $\mu'_{\infty}\ne 0$ can be easily obtained
through a shift of the variable $\eta$, so that the same results
hold in full generality. The generalization of the distribution
given in \Eqref{eq-p-eta-delta} reads
\begin{equation}
p(\eta) = D\, \delta(\eta+\mu'_{\infty}) + (1-D)\, c(\eta)
\end{equation} 
with the support of $c(\eta)$ limited to $[-\mu'_{\infty} ,\infty)$.

In the following subsection, we check that the distribution
\Eqref{eq-p-eta-delta}, obtained through necessary conditions,
indeed leads to a finite $\alpha_{\mathrm{c}}$, and thus to the possibility
of the absence of the glass transition.

\subsection{Distributions with a discrete mass at the minimal energy}

We now wish to show that if one starts from the distribution $p(\eta)$
given in \Eqref{eq-p-eta-delta}, the resulting function $\zeta(\beta)$
converges to a finite limit $\alpha_{\mathrm{c}}$.
This result is not obvious from the previous subsection, where we used
necessary conditions only, and did not study the behavior of the
term $\beta \mu'(\beta)$ in $\zeta(\beta)$.
Starting from \Eqref{eq-p-eta-delta}, $\mu(\beta)$ can be expressed as
\begin{equation} \label{eq-mubeta-D}
\mu(\beta) = \ln \big( D+(1-D)\, I(\beta) \big),
\end{equation}
where we have defined
\begin{equation} \label{def-Ibeta}
I(\beta) = \int_0^{\infty} c(\eta)\, e^{-\beta\eta} d\eta.
\end{equation}
The derivative $\mu'(\beta)$ is then given by
\begin{equation} \label{eq-muprime-D}
\mu'(\beta) = \frac{(1-D)\, I'(\beta)}{D+(1-D)\, I(\beta)}.
\end{equation}
As shown in \Eqref{eq-integ-split}, an integral of the form of $I(\beta)$
converges to zero when $\beta \to \infty$ for any distribution $c(\eta)$
which has no Dirac delta at $\eta=0$.
We thus have $\mu'(\beta) \sim (1-D)\, I'(\beta)/D$ for $\beta \to \infty$.
The evaluation of $\beta \mu'(\beta)$, which is needed to compute
$\zeta(\beta)$, thus boils down to that of $\beta I'(\beta)$.

The behavior of $\beta I'(\beta)$ can be evaluated by performing
an integration by part, yielding
\begin{equation}
\beta I'(\beta) = - \int_0^{\infty} \frac{d}{d\eta} \big( \eta\, c(\eta)\big)
\, e^{-\beta\eta} d\eta.
\end{equation}
Following the same arguments as for $I(\beta)$, one can then show that
$\beta I'(\beta)$ goes to zero when $\beta \to \infty$,
and so does $\beta\mu'(\beta)$.
Hence from \Eqref{eq:def-zeta}, $\zeta(\beta)$ converges to
the finite limit $\alpha_{\mathrm{c}}=-\ln D>0$ when $\beta \to \infty$.

\begin{figure}
\centerline{ \includegraphics[width=6cm]{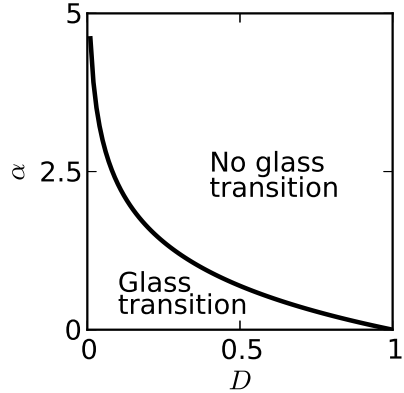} }
\caption{\label{fig:nega} Phase diagram in the $(\alpha,D)$ plane, for the distribution
$p(\eta)$ defined in \Eqref{eq-p-eta-delta}. The line $\alpha_{\mathrm{c}}=-\ln D$ separates regions
where the glass transition exists ($\alpha<\alpha_{\mathrm{c}}$) from regions where
no glass transition occurs.}
\end{figure}

In conclusion, for a distribution $p(\eta)$ of the form
\Eqref{eq-p-eta-delta}, we have shown that the glass transition disappears for
$\alpha > \alpha_{\mathrm{c}}$, with
\begin{equation} 
\alpha_{\mathrm{c}}=-\ln D,
\end{equation}
 as sketched in \figref{fig:nega}.

\subsection{Behavior of the glass transition close to the onset threshold} \label{sect-onset-glass}

We explore some interesting consequences of the above results.
For $\alpha<\alpha_{\mathrm{c}}$, the glass transition exists, but the temperature range
of the glassy phase is expected to shrink when $\alpha \to \alpha_{\mathrm{c}}$.
Let us make the argument quantitative.
The glass transition temperature is determined from the relation
$\zeta(\beta_{\mathrm{g}}) = \alpha$ [see \Eqref{eq-betag-alt}].
For $\alpha$ close to (and smaller than)
$\alpha_{\mathrm{c}}$, $\beta_{\mathrm{g}}$ is thus determined by the asymptotic
behavior of $\zeta(\beta)$ for large $\beta$.

To make concrete calculations, we assume that $c(\eta) \sim c_0 \eta^{\nu-1}$
when $\eta \to 0$ ($\nu >0$). Using the change of variables $u=\beta\eta$
in the integral defining $I(\beta)$, see \Eqref{def-Ibeta},
one finds for $\beta \to \infty$ that
\begin{equation} \label{eq-asympt-Ibeta}
I(\beta) \sim \frac{1}{\beta} \int_0^{\infty} c_0 \left(\frac{u}{\beta}\right)^{\nu-1} e^{-u} du = \frac{\Gamma(\nu) c_0}{\beta^{\nu}}.
\end{equation}
$I'(\beta)$ can be computed in the same way, simply replacing $c(\eta)$
by $-\eta c(\eta)$, which amounts to replacing $\nu$ by $\nu+1$
and $c_0$ by $-c_0$.
One then finds $I'(\beta) \sim -\Gamma(\nu+1) c_0/\beta^{\nu+1}$.
Taking into account Eqs.~(\ref{eq-mubeta-D}), (\ref{eq-muprime-D}) and (\ref{eq-asympt-Ibeta}), we obtain
\begin{equation}
-\ln D - \zeta(\beta) \sim \frac{(1-D)(\nu+1) \Gamma(\nu) c_0}{D\, \beta^{\nu}}.
\end{equation}
Using $\zeta(\beta_{\mathrm{g}}) = \alpha$, we get
\begin{equation} \label{eq-betag}
-\ln D - \alpha \sim \frac{(1-D)(\nu+1) \Gamma(\nu) c_0}{D\, \beta_{\mathrm{g}}^{\nu}}.
\end{equation}
At this stage, two different viewpoints can be adopted, namely either
considering $\alpha$ as the control parameter for a fixed $D$, or
considering $D$ as the control parameter for a fixed $\alpha$.
We first fix $D$, and use $\alpha$ as control parameter.
In this case, the glass transition occurs for $\alpha = \alpha_{\mathrm{c}} \equiv -\ln D$. From \Eqref{eq-betag}, we obtain for the glass transition temperature
$T_{\mathrm{g}} \equiv \beta_{\mathrm{g}}^{-1}$
\begin{equation}
T_g \sim A\, (\alpha_{\mathrm{c}} - \alpha)^{1/\nu}, \qquad \alpha \to \alpha_{\mathrm{c}}^-,
\end{equation}
with $A=[(1-D)(\nu+1) \Gamma(\nu) c_0/D]^{-1/\nu}$.

Alternatively, using $D$ as control parameter for a fixed $\alpha$,
the transition exists only for $D<D_c\equiv e^{-\alpha}$.
\Eqref{eq-betag} then leads to
\begin{equation}
T_g \sim \tilde{A}\, (D_c - D)^{1/\nu}, \qquad D \to D_c^-,
\end{equation}
with $\tilde{A}=[(1-e^{-\alpha})(\nu+1) \Gamma(\nu) c_0]^{-1/\nu}$.

\section{Discussion}
\label{sect-discus}

In summary, we have shown that by tayloring the distribution $p(\eta)$
of local energies (and thus modifying the global energy distribution
$\mathcal{P}(E)$), the glass transition can be avoided in some parameter
regimes, so that
the presence of uncorrelated random energy levels is not a
sufficient condition for the emergence of a glass transition.
Reversing the perspective, one could also interpret the onset
of a glass transition when varying either $\alpha$ or the weight $D$
of the Dirac mass, as a kind of critical phenomenon, the order
parameter of this transition being the glass transition itself.
We have seen in Sect.~\ref{sect-onset-glass} that the
glass transition temperature indeed behaves as a power law
close to threshold. However, the corresponding exponent
is non-universal, as it depends on the behavior of the distribution
$p(\eta)$ close to the lower bound of its support.

These results further suggest an interesting analogy with another type
of critical phenomenon. We have seen that if we build the
random energies $E_k$ of the model by adding up random positive terms
$\eta_{k,i}$, such that each term is either equal to zero with probability
$D$, or drawn from a continuous distribution $c(\eta)$ with probability
$1-D$, the glass transition disappears if the fraction of zero terms
exceeds some threshold $D_c$.
This situation is reminiscent of the dilute Ising model, where
the coupling constants between neighboring sites are randomly set to zero
with a given probability. Above a critical fraction of zero couplings,
the transition disappears \cite{Aizenman,Parisi,Vicari}.
The mechanism at play in the dilute Ising model is however different,
as it is related to the percolation of the bonds with nonzero couplings,
and thus has a geometric interpretation.

Besides, it is interesting to study the properties of the large deviation
function $\phi(\epsilon)$, which is related to the microcanonical
entropy $s_\mathrm{m}(\epsilon)$ through
$s_\mathrm{m}(\epsilon)=\alpha-\phi(\epsilon)$ [see \Eqref{eq:ztyp}].
Using \Eqref{eq:mu-beta-phi} and the properties of the Legendre transform,
one can show that $\phi(\epsilon_{\mathrm{m}}(\beta))=\zeta(\beta)$.
Denoting as $\eta_{\min}$ the lower bound of the support of the distribution
$p(\eta)$, one has $\epsilon_{\mathrm{m}}(\beta) \to \eta_{\min}$ when
$\beta \to \infty$ (note that $\eta_{\min}$ may be equal to $-\infty$).
Hence if $\zeta(\beta) \to \infty$ for $\beta \to \infty$,
$\phi(\epsilon)$ also diverges for $\epsilon \to \eta_{\min}$.
In this case, the equation $\phi(\ecut)=\alpha$ always has a solution
$\ecut > \eta_{\min}$, assuming that $\phi(\epsilon)$ is continuous.
Furthermore, as $\phi(\epsilon)$ is regular at $\epsilon=\ecut$,
its derivative is finite, and so does the glass transition temperature
$\beta_{\mathrm{g}}=\phi(\ecut)$ --see Eqs.~(\ref{eq:phiprime-beta}) and
(\ref{eq:epsm-ecut}).

In constrast, if $\zeta(\beta)$ converges to a finite limit,
$\phi(\epsilon)$ also has a finite limit, equal to $\alpha_{\mathrm{c}}$,
for $\epsilon \to \eta_{\min}$.
In this case, as $\phi'(\epsilon_{\mathrm{m}})=-\beta$ (see \Eqref{eq:phiprime-beta}), 
the large deviation function necessarily has an infinite negative slope
for $\epsilon \to \eta_{\min}$.
Hence the absence of glass transition in this case does not have the same
origin as in non-disordered systems. For such systems, there is no cut-off
$\ecut$, and energies down to $\eta_{\min}$ can be explored.
In the presence of disorder, the glass transition can be avoided
only by tayloring $\phi(\epsilon)$ so that the equation determining the cut-off
$\ecut$ has no solution $\ecut>\eta_{\min}$.

Let us finally mention that the glass transition in the REM also has
applications in other fields, like the empirical estimation of moments
in statistical signal processing.
It has been recently emphasized that the so-called ``linearization effect''
in multifractal analysis, occuring when empirically determined moments
$S_M(q)=M^{-1}\sum_{i=1}^M x_i^q$ significantly
depart from the theoretical ones $\langle x^q \rangle$,
can be interpreted as an analog of the glass transition in the REM
\cite{Angeletti11,MuzyBacryKozhemiak2006, MuzyBacryKozhemiak2008, BacryGlotterHoffmannMuzy2010}, the inverse temperature $\beta$ being mapped
onto the moment order $q$.
Similar effects also occur even when considering uncorrelated signals,
as long as the marginal distribution of the signal is sufficiently broad,
but with finite moments (for instance a lognormal distribution)
\cite{Angeletti12}.
Along this line of thought, the present version of the REM can be
seen as the analog of the moment estimator of variables built
as products of a large number of independent random variables
(which is a possible way to build broadly distributed variables
with finite moments).
The present study then shows that in most cases, this moment estimator
will also present a linearization effect, and depart from the theoretical
moments for $q$ above some threshold $q^*$.
However, if the underlying distribution of the variables presents a Dirac
mass at its upper bound, this Dirac mass generates a linear branch in the logarithm of the theoretical moments which can mask the empirical linearization effect.

\hspace*{5mm}

\end{document}